\documentclass[runningheads]{llncs}
\usepackage{cite}
\usepackage{color}
\usepackage{multirow}
\usepackage{booktabs,multirow,array}
\usepackage[table]{xcolor} % load both xcolor and colortbl
\usepackage{ltxcmds}
\usepackage{footnote}
\usepackage[center]{caption}
\usepackage{array}

\usepackage[table]{xcolor}
\usepackage{subfigure}
\usepackage{lipsum}

\usepackage{soul} %%% Shan: for highlighting 

\usepackage{amssymb} %this is for checkmark symbol
\usepackage{pdflscape}
\usepackage{afterpage}

\usepackage{pifont}% http://ctan.org/pkg/pifont
\graphicspath{{Figures/}}

\usepackage{graphicx}
\begin{document}
\title{Challenges and Opportunities of Blockchain for Cyber Threat Intelligence Sharing}
\titlerunning{Blockchain for Cyber Threat Intelligence Sharing}
\author{Kealan Dunnett\inst{1} \and
Shantanu Pal \inst{1} \and
Zahra Jadidi \inst{2}
}
\authorrunning{K. Dunnett et al.}
\institute{School of Computer Science, Queensland University of Technology, \\ Brisbane, QLD 4000, Australia \and
School of Information and Communication Technology, Griffith University, \\Gold Coast Campus, QLD 4222, Australia\\
\email{kealan.dunnett@connect.qut.edu.au,  shantanu.pal@qut.edu.au,
z.jadidi@griffith.edu.au}
}
\maketitle              % typeset the header of the contribution

\begin{abstract}
% The emergence of Internet of Thing (IoT) technology has caused a drastic evolution in the threat landscape. As a result, Organisations have had to find new ways to better manage the risks associated with their infrastructure. In response to this, a significant amount of research focused on developing efficient Cyber Threat Intelligence (CTI) sharing platforms can be observed in the current literature. This work subsequently seeks to evaluate how blockchain technology can facilitate CTI sharing, as current approaches mostly utilise centralised architectures. To determine the role of blockchain-based sharing moving forward, we present a number of general CTI sharing challenges and discuss how blockchain can bring opportunities to address these challenges. Finally we discuss a number of relevant works and deliver a number of future research questions. 

The emergence of the Internet of Things (IoT) technology has caused a powerful transition in the cyber threat landscape. As a result, organisations have had to find new ways to better manage the risks associated with their infrastructure. In response, a significant amount of research has focused on developing efficient Cyber Threat Intelligence (CTI) sharing platforms. However, most existing solutions are highly centralised and do not provide a way to exchange information in a distributed way. In this chapter, we subsequently seek to evaluate how blockchain technology can be used to address a number of limitations present in existing CTI sharing platforms. To determine the role of blockchain-based sharing moving forward, we present a number of general CTI sharing challenges, and discuss how blockchain can bring opportunities to address these challenges in a secure and efficient manner. Finally, we discuss a list of relevant works and note some unique future research questions. 

\keywords{Blockchain  \and Cyber threat intelligence \and Security \and Information sharing.}
\end{abstract}

%%% New Section 
%%% New Section
\section{Introduction}
\label{introduction}

%{\color{red} Shan: make introduction another half a page longer} {\color{blue} KD: Done}

Each year the threat landscape continues to evolve in both the types of cyber-attacks and the methods used to commit them~\cite{lamssaggad2021survey}. Organisations have subsequently had to find ways to manage the increased risk associated with the infrastructure they depend on to operate. As a result, several Cyber Security Risk Management (CSRM) frameworks have been developed to define a more concrete framework to better manage this risk~\cite{kure2022integrated}. However, with the emergence of the Internet of Things (IoT) technology~\cite{corallo2022cybersecurity}, smart portable sensors and their resource-constrained nature, the threat landscape has recently grown at a rate that makes the traditional CSRM task challenging~\cite{rabehaja2019design}~\cite{pal2017design}. To minimise cyber threats, organisations continue to develop methods focused on gathering threat-based information specific to them. Towards this, Cyber Threat Intelligence (CTI) is a concept that describes the collection and analysis of threat information by an organisation. The emergence of CTI in recent years has seen its integration into traditional CSRM frameworks become a effective threat mitigation strategy \cite{kure2019cyber}. 

The SANS institute is a US based organisation that conducts a yearly CTI survey across industry. The primary aim of this survey is to understand the current state of CTI use within industry. In their 2021 survey, a significant milestone was reported, 100\% of surveyed organisations indicated that they either currently do or have plans to use CTI in some way \cite{brown20212021}. When this figure is contrasted with the 75\% reported only four year earlier in 2017, it is clear that CTI will continue to play a critical role in threat mitigation within industry moving forward.  

Sharing CTI cooperatively between organisations can be highlighted as a mutually beneficial process for all participating organisations \cite{johnson2016guide}. However, in practise CTI sharing is challenging due to the variety of ways threats can affect the components that make up an organisations infrastructure (e.g., Storage, Networks and Communication). For example, the man-in-the-middle attack, eavesdropping attack, phishing and spear-phishing attacks, etc~\cite{pal2020security}. 

%Above Paragraph
%{\color{green} Is CTI sharing challenging because of the types of attacks? This paragraph is not clear.}
%{\color{blue} (KD: Done)}
%{\color{red} (Shan: which information, always make it clear)} 
%{\color{blue}(KD: Done, will try to look for more of these)}

In recent years Vendor-created/Open-source threat intelligence sharing platforms, have become a popular choice within industry. These platforms provide organisations with an environment where they can share and consume CTI in either a fully or semi automated way. During their 2021 survey the SANS institute noted that these types of sharing platforms saw a 3\% increase in use compared to 2020 \cite{brown20212021}. Moreover, it was also reported that more traditional sharing mechanisms (e.g., emails and briefs) saw a 7.8\% decrease in use compared to 2020. 

We argue that while this trend towards either fully or semi automated threat intelligence sharing is positive, a number of key challenges (e.g., Produce Consumer Imbalance, Data Validity) are currently prevalent in this space, as highlighted by existing literature \cite{wagner2019cyber}. Furthermore, we also seek to provide a unique insight into how privacy, trust and accountability define a seemingly paradoxical relationship. As well as discussing several general CTI sharing challenges, we also seek to demonstrate that using a decentralised platform for CTI sharing between organisations in a trustless manner has tremendous promise. 

Towards this, blockchain is a promising technology. Blockchain is a tamper-proof, decentralised, and immutable storage of digital information that is impossible to change~\cite{nakamoto2008bitcoin}. Therefore, blockchain can provide a strong and effective solution for securing CTI in networked ledgers, a series of blocks that are cryptographically linked, and facilitates secure dissemination between organisations. However, blockchain-based CTI sharing solutions are lacking in the present literature. A few proposals, e.g., \cite{badsha2020blocynfo} \cite{homan2019new} \cite{gong2020blocis}, integrate blockchain for CTI sharing, but a comprehensive solution which addresses all of the discussed challenges is currently lacking.

In this chapter, we evaluate a number of present CTI sharing challenges and discuss how blockchain can bring opportunities to address these challenges. Thus, the major contribution of this chapter is to provide a list of challenges associated with CTI sharing and deliver a list of opportunities present within the blockchain space for future research.

The remainder of the Chapter is organised as follows. In Section~\ref{overvicew-blockchain-cti}, we present a brief overview of blockchain and CTI. In Section~\ref{bc-cti}, we present a simplified blockchain-based CTI sharing model from the current literature to demonstrate how blockchain can facilitate sharing. In Section~\ref{challenges}, we discuss the a number of challenges associated with CTI sharing. In Section~\ref{opportunities}, we present a number of opportunities that highlight the applicability of blockchain-based models in the CTI sharing space based on current ideas presented in the literature. In Section~\ref{related-work}, we present a brief discussion the related work within the literature. To concluded, in Section~\ref{conclusion-future-work} we summarise the work presented in this Chapter and discuss future research directions. 

%{\color{red} Shan:
%To do:
%1. Complete the abstract {\color{blue}(KD: Done)}\\
%2. Say in this Chapter, not paper/article {\color{blue}(KD: Done)}\\
%3. Add appropriate references, I see it is missing \\
%4. You use information and data interchangeably, use either. I would go with  information {\color{blue}(KD: Done)}
%} 

%%%
\section{Overview of Blockchain and CTI}
\label{overvicew-blockchain-cti}
%In this section, we present a brief overview of blockchain and CTI. 
Before discussing blockchain-based CTI sharing in detail, we present a brief overview of blockchain and CTI in this section.

\subsection{Blockchain}
\label{blockchain}
Blockchain is a distributed digital ledger for storing electronic records \cite{nakamoto2008bitcoin}. In other words, blockchain can be seen as a network of computers that store transactions (and therefore the data) across multiple computers. These computers are considered a node in the blockchain. The data entered in a particular interval in the chain is known as a \textit{block}. Each block is identified using a unique identifier is called a \textit{hash}. Each block contains the hash of the previous block. A hash is the output of a unique cryptographic function that takes as input a arbitrary amount amount of data and generates a fixed-size output, the hash. Significantly, this is a one-way function and it is impossible to reserve the computation~\cite{wang2018research}. 

In blockchain, when a transaction is first equested, it is authenticated using cryptographic keys (public and private keys). Then a block containing that transaction is created and sent to the entire network. Once the transaction is agreed between the nodes in the network, it is approved (i.e., authorised) before the block is added to the chain. This is done by a mechanism called \textit{consensus}, where the majority of nodes agree with the transaction. Note that nodes must perform a complex mathematical problem to validate a transaction. This is known as \textit{mining}, and the participanting nodes are referred to as the \textit{miners}. Commonly, the mining task in called \textit{Proof of Work (PoW)}. A cryptocurrency reward is given to the miner who first solves the mathematical problem (i.e., the PoW) and validates a block. After this, the block is added to the existing chain, and all the nodes in the network are updated with this information~\cite{wang2019survey}. 

Therefore, blockchain provides a framework in which nodes can maintain an immutable ledger of data. In Fig.~\ref{fig:x blockchainExample}, we illustrate how immutability is created in blockchain by linking successive blocks together using cryptographic hashing functions. Currently PoW is the most widely used mechanism for mining. However, PoW requires a substantial computing power and therefore uses considerable amounts of energy, a notable drawbacks of PoW. To solve this issue, another mining mechanism, \textit{Proof-of-Stake (PoS)} is becoming popular. PoS provides faster transactions and uses less energy during mining~\cite{lepore2020survey}. Some significant properties of blockchain are outlined as follows \cite{dai2019blockchain}:

%\lipsum
\begin{figure*}
\centering
\includegraphics[scale=0.58]{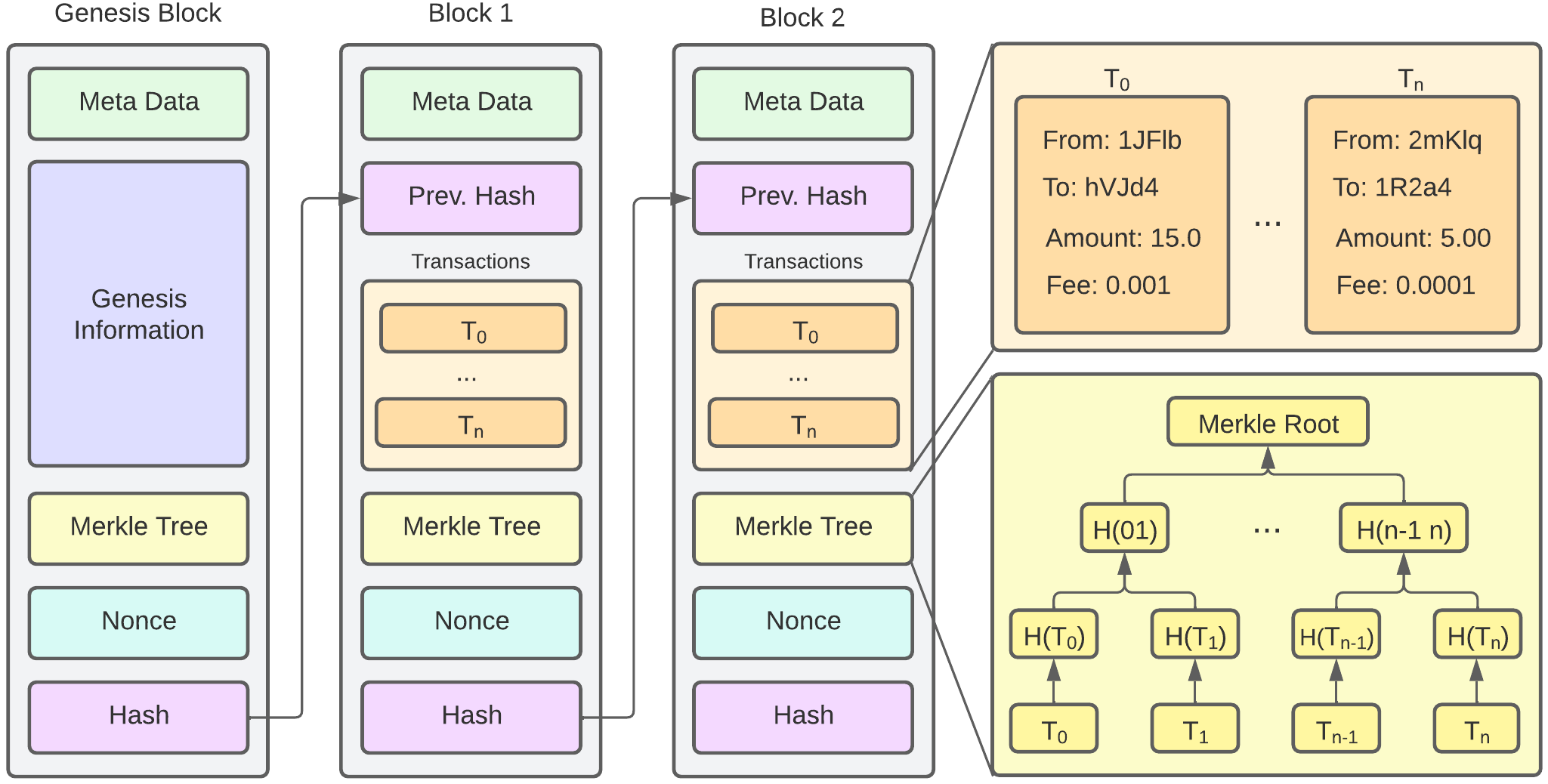}
\caption{An illustration of blockchain immutability created by hashing}
\label{fig:x blockchainExample}
\end{figure*}

\begin{itemize}
\item \textit{Decentralised:} Does not require a single central authority to validate transactions rather a consensus algorithm is used. As a result, Blockchain does not suffer from a single point of failure.  

\item \textit{Immutability:} Once a block is added to the chain it is almost impossible to delete or change the data. This provides security to the stored data.

\item \textit{Anonymity:} Provides nodes with the ability to participate without disclosing their identity. 

\item \textit{Trustless:} Nodes also do not have to have pre-established trust with each other to transact, with all transactions documented on the ledger to ensure transparency.

\item \textit{Auditability:} At any point in time an existing transaction on the chain can be validated. To ensure a transaction has not been changed or altered the proceeding blocks hashes can be checked. 

\item \textit{Transparency:} Every transaction that takes place is stored on the blockchain and therefore is visible to the every node in the network. 

\item \textit{Use of Smart Contracts:} Transaction in the blockchain can be automated with smart contracts. It is a computer code that facilitates and verifies the nodes' agreements and therefore increases the computational efficiencies.

\end{itemize}

%%%
\subsection{CTI}
\label{CTI}
Cyber Threat Intelligence refers to a collection of evidence-based knowledge about cyber threats. This knowledge can be compromised of a variety of information including Indicators of Compromise (IoC), attackers’ motivations, intentions, characteristics, attack vectors, as well as their Techniques, Tactics, and Procedures (TTPs) \cite{moubarak2021dissemination}. CTI can also consist of actionable advice to detect, prevent, and mitigate the impact of attacks. It can also be obtained from a variety of sources, including anti-virus programs, open-source intelligence (OSINT), Intrusion Detection Systems (IDSs), human intelligence, malware analysis, code repositories, and CTI sharing platforms. CTI can be categorised into following four types: (i) strategic, (ii) operational, (iii) tactical, and (iv) technical. A brief description for each type follows:

\begin{itemize}
    \item \textit{Strategic} CTI provides a high-level overview of the threat landscape in terms of past, current, and future trends. This type of CTI is often presented in plain language and is focused on improving situational awareness and presenting business risks. The intended audience is senior, lay-person decision makers in an organisation.
    
    \item \textit{Operational} CTI refers to information about the nature and motivations of potential upcoming attacks against an organisation, that can be used to formulate targeted prevention strategies and prevent future incidents.
    
    \item \textit{Tactical} CTI relates to TTPs and IoCs, that are useful to identify specific attack vectors and vulnerabilities for the purposes of proactively updating signature-based defences against known threats.
    
    \item \textit{Technical} CTI consists of technical information often found on threat intelligence feeds about malware and adversarial campaigns, including information about an attacker’s assets, attack vectors employed, Command and Control domains used, and types of vulnerabilities exploited.
\end{itemize}

CTI deals with the collection and analysis of evidence-based knowledge about existing or potential threats that can be used to inform decision making. The aim of CTI is to aggregate a number of unstructured data sources (e.g., network logs and software signatures) and create structured intelligence that details a threat \cite{kure2019cyber}.

As noted in Section \ref{introduction}, traditional CTI sharing systems lack the ability to share this intelligence effectively. Several of the major challenges that these systems have yet to overcome are -- the producer consumer imbalance, data validity, legal and regulatory factors, and sharing intelligent intelligence. Consequently, a number of recently proposed CTI sharing platforms have integrated blockchain into their design to try and provide novel solutions to these challenges.

%Above paragraph
%{\color{red} (Shan: tools or systems would better fit?)} {\color{blue} (KD: Systems)}

\section{Blockchain-Based CTI Sharing}
\label{bc-cti}

\begin{figure}[t]
\centering
\includegraphics[scale=0.75]{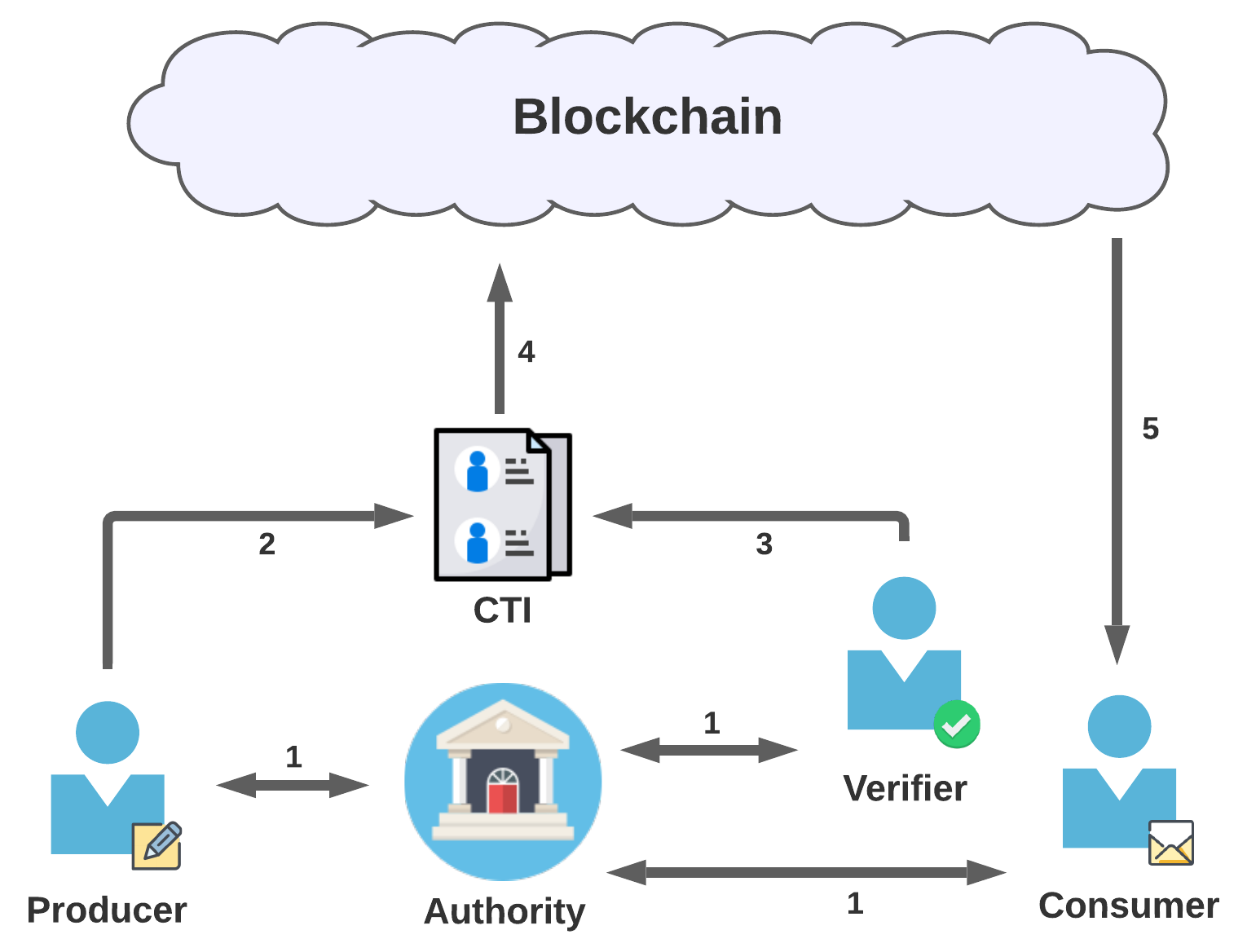}
\caption{A typical blockchain-based CTI sharing framework.}
\label{fig:x shareModel}
\end{figure}

Significant diversity exists in the blockchain-based CTI sharing space. These models utilise specific blockchain characteristics and cryptographic constructs in a variety of ways to facilitate sharing. In Fig.~\ref{fig:x shareModel}, we illustrate a simple sharing framework which exemplifies how blockchain can be applied to CTI sharing \cite{nguyen2021blockchain}. This model is composed of the following components.

\begin{itemize}
    \item \textit{Consumers:} Users who consume shared CTI information. Make decisions about which intelligence they consume based on the relevance to their physical infrastructure or business case.  

    \item \textit{Producers:} Users who produce CTI based on internal information that can be linked to an existing or new threat. This CTI is then shared with an individual, group or publicly, based on sensitivity of the intelligence using blockchain.

    \item \textit{Verifiers:} Users who validate shared CTI to ensure it meets sharing standards (e.g., Complies with STIX format), is not duplicated intelligence that has already been shared, and or maliciously contains fake information. The results of this user’s analysis either directly impacts the addition of CTI to the blockchain or is added with the given CTI as a report to inform consumer decisions. 

    \item \textit{Authority:} Users who verify the identity of other users before they participate in sharing. This authentication creates trust between users who produce and consume intelligence as they can be sure that only authenticated users are able to do so.

    \item \textit{Blockchain:} It is used to provide a distributed ledger of CTI information (e.g., Hyperledger, Ethereum, EOS, etc).
    
    \item \textit{CTI Smart Contract:} Self managed code that is executed by the blockchain to manage the verification of shared CTI. This contract is made up of a Inter Planetary File System reference to the shared CTI and a verification status. 
\end{itemize}

\textit{Note: Users can be any combination of the above roles and subsequently are not restricted to one role.} \\

As shown in Fig.~\ref{fig:x shareModel}, the process of communication among the various components of the framework follows these steps. 

    \begin{itemize}
        \item \textit{Step 1:} All stakeholders prove their identity to a trusted Authority. Proof-of-identity can consist of the exchange of information like government credentials (e.g., drivers licence or passport), ownership of third party certificates or industry accreditation. 
        \item \textit{Step 2:} Producer generates CTI and adds it to the blockchain for verification.
        \item \textit{Step 3:} Verifier determines the credibility of the CTI based on a set of standards agreed upon by the network.
        \item \textit{Step 4:} CTI that is determined to be valid in \textit{Step 3}, is added to the blockchain.
        \item \textit{Step 5:} Consumers access CTI that has been added to the blockchain. 
        %{\color{green} How do we verify the identify of the consumer?} {\color{blue} (KD: Will add a line here)}
        
    \end{itemize}

%{\color{red} (Shan: Two three lines need here to summarize the section - what is your findings)} {\color{blue} (KD: Done)}

The simplified sharing model presented in Fig.~\ref{fig:x shareModel} demonstrates how blockchain can be used to facilitate CTI sharing at a basic level. Moreover, when the properties of blockchain discussed in Section \ref{blockchain} are considered in the context of CTI sharing, the advantages that blockchain-based sharing models have over traditional centralised approaches can be highlighted.   

%Above
%{\color{blue} (KD: Not to sure how to reference sections properly)}

%%% New Section
\section{Challenges}
\label{challenges}

\begin{figure*}
\centering
\includegraphics[scale=0.49]{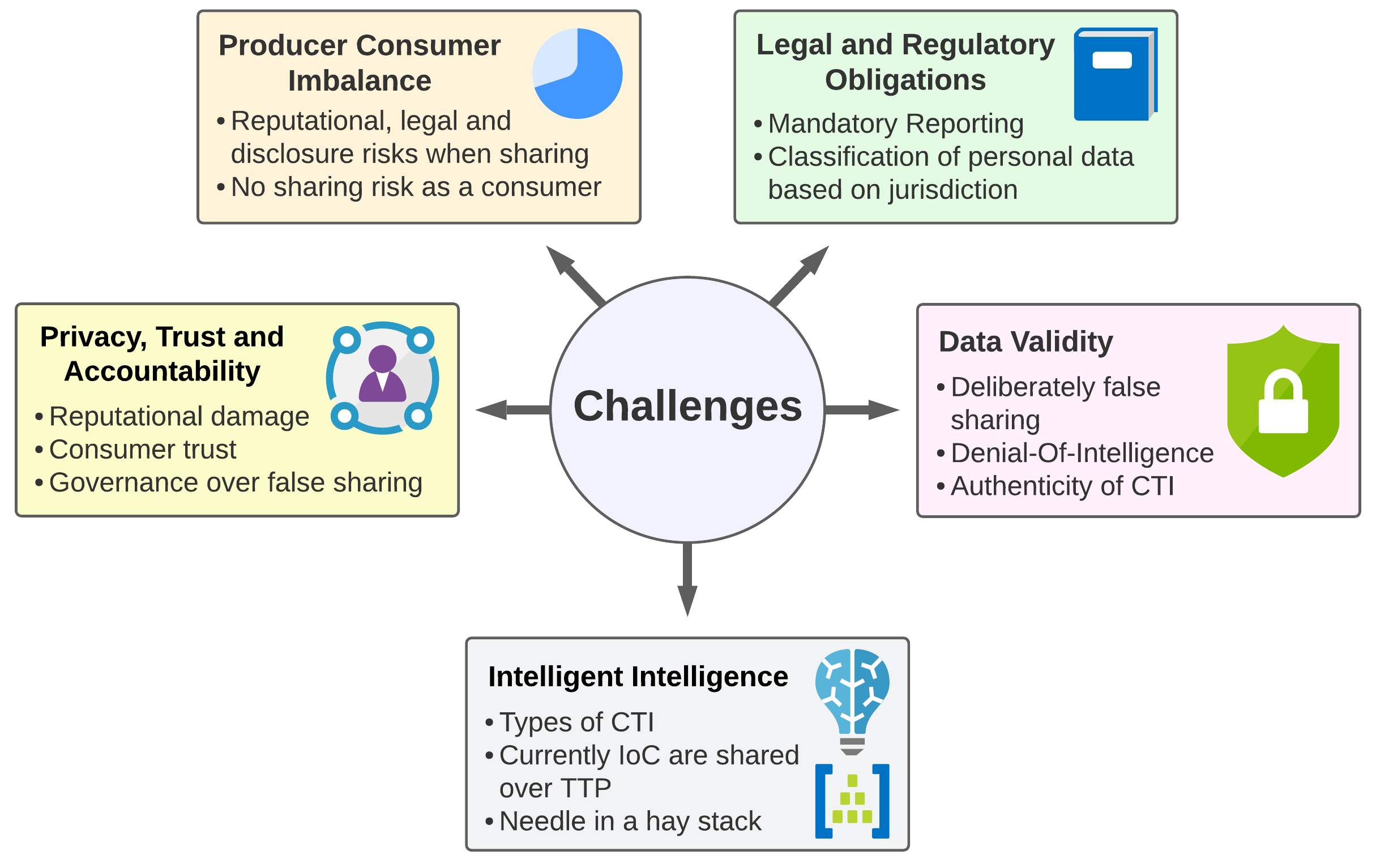}
\caption{General CTI sharing challenges.}
\label{fig:x challenges}
\end{figure*}

Traditional CTI sharing frameworks (e.g., MISP, OpenCTI and ISACs) have a wide range of challenges that are documented in the literature \cite{wagner2019cyber} \cite{tounsi2018survey}. In this section, we focus on a subset of these general CTI sharing challenges (cf. Fig.~\ref{fig:x challenges}).

%%%
\subsection{Producer Consumer Imbalance}
\label{pc-imbalance}
Stakeholder who participate in CTI sharing as either a producer or consumer (cf. Section \ref{bc-cti}) must consider the risks and benefits associated with doing so. In the case where a producer shares intelligence, a number of reputational and or monetary risks are prevalent. For example, sharing intelligence that indicates an organisation has been the victim of a ransomware attack, could cause stock prices to fall or new customers to choose a competitor. Some of the potential risks are listed below \cite{wagner2019cyber} \cite{tounsi2018survey};

\begin{itemize}
    \item\textit{Consumer Distrust:} Potential consumers might feel that a reported cyber incident means that the organisation is vulnerable. As a result, existing customers may decide to use the services of a competitor that has not reported an incident.
    
    \item\textit{Competitor Advantage:} Competitors become aware of potential vulnerabilities that might affect them without being directly affected by it. This allows them to implement mitigation strategies for the same vulnerability at a reduced resource cost.
    
    \item\textit{Revealing Trade Secretes:} Information about hardware, software or services an organisation use might be revealed. %{\color{red} (Shan: incident report, but your heading said trade secretes?)} {\color{blue} (KD: Done)}
    
    %{\color{green} (Zahra: Why did you repeat this sentence? Should we delete this second one?)}{\color{blue} (KD: Removed)}
    
\end{itemize}

Apart from being able to consume CTI themselves, producers do not gain any direct benefits from sharing. As a consequence, without implementation of a reward-based system as part of a sharing platform, the process of sharing CTI can be considered a common good service. On the other hand, consumers assume almost no risk when consuming CTI. Even in the case where the consumption of specific CTI is attributable to an organisation, this action alone is not likely to result in the same reputational or monetary consequences associated with sharing. Given that organisations that consume CTI can implement mitigation strategies against vulnerabilities before they can be acted on, we propose that the following benefits that could be gained;

\begin{itemize}
    \item\textit{Increased Service Quality:} Increase service up time provides existing customers with a better service quality. This could result in a higher customer retention rate. As a result of providing existing customers with a more consistent service, an organisation might gain a reputation for providing services with low downtime. 
    
    \item{\textit{Reduced Negative Publicity:}} In the case where an organisation successfully implements a mitigation strategy to fix a shared vulnerability, the potential for negative publicity due to a successful attack is removed.
    
    \item\textit{New Customers:} In the case where an organisation has suffered from a number of cyber incidents (e.g., DDoS Attacks, Privacy Leak), it could be predicted that dissatisfied customers could seek an alternative service. Moreover, if a competing organisation providing and analogous service that has not suffered from these same incidents due to the consumption of CTI, it could be predicted that this organisation could gain additional customers.   
\end{itemize}

From the above discussion it is clear that the risks and benefits associated with the producer and consumer role are not equal. This inequity, consequently creates an imbalance. If this imbalance is not addressed as part of a sharing platforms design, organisations can be observed to exhibit \textit{free-riding} behavior \cite{wagner2019cyber}. In this case, \textit{free-riding} behavior can be defined as a deliberate lack of participation by organisations who could share valuable CTI, however choose not to. If a large portion of organisations deliberately behave in this way, the productivity of a sharing platform is affected in two major ways \cite{ring2014threat};

\begin{enumerate}
    \item Not sharing removes the ability of other organisations to mitigate against the same incident. When CTI is shared, it is possible for other organisations to put in place mitigation strategies (e.g., Firewall rules) to ensure they are not susceptible to attacks which have a similar profile or share common characteristics. In the case of \textit{free-riding}, this is not possible.  
    
    \item \textit{Non-free-riding} organisations might stop or reduce the amount of intelligence they share due to a lack of consumable CTI from others. As noted above, producers assume a number of risks when they participate in sharing. However, if part of a productive platform where a large volume of valuable intelligence is shared, this risk compared to the benefit gained by consuming other intelligence makes sharing more attractive. Consequently, a large portion of \textit{free-riding} organisations has the potential to impact the sharing behaviours of others.
\end{enumerate}

\subsection{Legal and Regulatory Obligations}
Organisations who participate in sharing have to follow the legal and regulatory obligations associated with the jurisdiction they are from. Survey \cite{wagner2019cyber}, highlights a number of legal and regulatory obligations that organisation in certain countries must meet. 

For example, in Germany Internet Protocol (IP) addresses are considered personal information and therefore any disclosure of CTI containing them must comply with German privacy laws \cite{nweke2020legal}. However, in the UK IP addresses are not considered personal information and therefore can be freely shared. In terms of CTI sharing, IP addresses are likely to be shared as an IoC and therefore organisations based in these different jurisdictions have to ensure they comply with the applicable laws. Moreover, countries like Belgium and Slovenia have mandatory sharing legislation \cite{wagner2019cyber}. This legislation requires organisations from these two countries to report any cyber incidents to a specific authority when they occur. If these organisations were also to participate in CTI sharing on top of this, in some cases the resources consumed to facilitate both of these independent sharing requirements could exceed those which are available. 

The above examples highlight that while theoretically CTI sharing is ubiquitous across the world, legal and regulatory obligations can pose a significant barrier. Given that legal and regulatory obligations are significantly diverse across the world, sharing platforms must ensure CTI can be shared in a pliable way.

\subsection{Data Validity}
\label{data-validity}
Threat hunting is defined by \cite{bynum2019cyber} and \cite{jadidi2021threat} as the proactive approach of seeking anomalous or malicious activity within an organisations cyber terrain. The process of performing this task, which if successful can result in the production of CTI, can be highly variable in nature. At the most basic level, threat hunting can simply consist of manual analysis of network or Windows logs. In contrast, \cite{arafune2022design} proposes a sophisticated threat hunting model which utilises machine learning to automatically generate threat intelligence based on data from a variety of sources. 

While these examples vary in their sophistication, they both share the common feature that the process of generating CTI is solely completed by the sharing organisation. As a result, it is possible for malicious organisations or individuals to intentionally generate and share false intelligence. We note that sharing false CTI has the potential to be utilised in several ways to either gain additional attack surfaces or to bury real CTI amongst fake intelligence going forward. Two examples of this are discussed below. %{\color{red} (Shan: Where are two examples, make it clear one and two...)} %{\color{blue}(KD: This is my observation, is it okay to make these kinds of statements? Or should we stick to what the literature already says?)}. {\color{red} (Shan: You can say so, then say we argue, or we observe, or we note ...)}

\subsubsection{Automatic Attack Feed Exploitation: }
Recent trends in CTI sharing have seen many notable developments towards automation, both in its generation (as discussed above) as well as in its consumption. For example, technologies such as Structured Threat Information Expression (STIX) and Trusted Automated eXchange of Intelligence Information (TAXII), has allowed many organisations to easily share and consume CTI in an automated way \cite{wagner2019cyber}. As CTI consumption becomes more automated, it could be feasible for threat actors to utilize this to create new attack surfaces. For example, intelligence structured using STIX can contain SNORT rules that consumers can automatically feed into their intrusion detection systems (IDS) \cite{tounsi2018survey}. Given certain conditions, we theorise that it could be plausible for an attacker to construct seemingly legitimate intelligence that causes a consuming organisations IDS to flag legitimate activity as malicious. This technique could be used in conjunction with an actual attack, to disguise malicious activity amongst legitimate traffic that is falsely flagged as suspicious. 

\subsubsection{Denial-of-Intelligence:}
As sharing platforms become more and more effective at allowing organisations to mitigate against threats, they themselves could become targets. Denial-of-Service (DoS) attacks have been around since the origin of the internet yet still remain highly effective in the present day. The main goal of a DoS attack is to simply make a particular computing resource unavailable \cite{alma2018DDoS}. The most common way that these types of attacks are committed, is by overwhelming a service with a large volume of bogus requests. We observe that a `Denial-of-X' style attack could be constructed to target CTI sharing platforms specifically. 

In this case, threat actors could develop Denial-of-Intelligence (DoI) attacks. This type of attack would seek to overwhelm a platform with a large amount of bogus intelligence. By flooding a sharing platform with a large amount volume of fake intelligence, threat actors could exploit a common vulnerability across multiple targets. The result of this would mean that while valid intelligence detailing the attack could be shared by the initial victim, it is buried amongst an overwhelmingly large volume of the false information. \\ %{\color{red} (Shan: Reference?)} {\color{blue}(KD: This was my observation so indicated this above)}

Both the above examples highlight that the ability to accurately determine the validity of shared CTI is a critical challenge that platforms must find novel ways overcome. Moreover, these examples also indicate that as the process of sharing becomes more automatic and widely used, data validity becomes more critical. 

%%%
\subsection{Intelligent Intelligence}
\label{intelligent-intelligence}
In Section \ref{CTI}, we discussed what CTI is and highlighted that it can be categorised into four main types: (i) strategic, (ii) operational, (iii) tactical, and (iv) technical. Each of these types of intelligence convey a narrative about a threat, however do so in diverse range of ways, specific to their intended recipients. For example, Technical CTI is made up of data that describes the physical attributes of an observed attack (e.g., IP address, MAC address, Malware Hash, etc), intended to be consumed by technical resources \cite{tounsi2018survey}. 

It is important to understand that these types of intelligence are highly variable in their sophistication. In this case, sophistication refers not just to the quality of the CTI itself, but how consumers are able to use it. Proposal \cite{moubarak2021dissemination} makes an important distinction between data, information, and intelligence, that highlights this variability. They are as follows: 

\begin{itemize}
\item \textit{Data:} Simple facts that can be made available in large volumes such as IP addresses, logs, hashes.

\item \textit{Information:} A collection of raw data together that shows suspicious activity.

\item \textit{Intelligence:} The process of analyzing and drawing meaningful conclusions that can be used by security professionals to define an intelligence-lead approach to decision making.
\end{itemize}

If the above criteria are applied to the categories of CTI discussed in Section \ref{CTI}, tactical, operational and strategic CTI could be classed as intelligent intelligence. On the other hand, technical intelligence (e.g. IoC) can only be classified as data/information intelligence, and subsequently cannot directly inform decision making. As a result, intelligence types can be grossly defined into high-level intelligence (e.g., TTP) and low-level intelligence (e.g. IoC). 

Currently, the majority of exchanged CTI can be classified as low-level intelligence  \cite{abu2018cyber} \cite{kure2019cyber} \cite{sauerwein2017threat}. Survey \cite{tounsi2018survey}, notes that over 250 million IoC are shared cumulatively across CTI sharing platforms every day, with this figure likely increasing in recent years. From the outset, this trend of sharing large volumes of technical intelligence may appear positive. However, when framed from the perspective of a consuming organisation, the quantity of available intelligence becomes an interpretability challenge analogous to the \textit{needle in a haystack} problem. 

\subsection{Privacy, Trust, and Accountability}
\label{privacy}
Privacy, Trust and Accountability, are three factors that any CTI sharing platform must balance to facilitate an environment conducive to share and consume CTI \cite{wagner2019cyber} \cite{allouche2021trade} \cite{pal2015extending}. The relationship between each of these factors and CTI sharing are discussed below; %{\color{red} (Shan: Just thinking, if you can draw a figure showing the relationship among trust, privacy, and accountability. Again, a random thought, do not waste more time on it)}

\subsubsection{Privacy}
can be defined as the ability or inability for a consuming organisations to associate some shared intelligence with the sharing organisations real identity. The literature consistently highlights reputational damage as a significant barrier that stops organisations from participating in CTI sharing \cite{tounsi2018survey} \cite{wagner2019cyber} \cite{abu2018cyber}. Given that reputational damage can result from sharing intelligence in an identifiable way, a degree of anonymity is required when sharing. 

\subsubsection{Trust}
can be defined as a consumers ability to trust the intelligence which they receive \cite{sauerwein2017threat}. Subsequently, a trust relationship between CTI producers and consumers is present in any sharing platform. In contrast to privacy, the parameters used to define the trust relationship between producers and consumers often require some link to the producer’s real identity. By linking at some level a producers real identity to the CTI they share, consumers have greater assurance that shared intelligence comes from an authoritative source~\cite{schaberreiter2019quantitative}.

\subsubsection{Accountability}
can be defined as the ability for a sharing platform to provide governance shared CTI. In this case, Governance refers to a sharing platforms ability to hold users who participate in false sharing responsible. Subsequently, the ability to hold users accountable for their actions insures that the integrity of shared intelligence can be maintained. Like trust, accountability is also dependent on being able to reveal a producers real identity given that they have made a malicious contribution~\cite{shin2020review} \cite{lueks2015revocable}.\\

From the above discussion, it can be hypothesised that privacy, trust, and accountability form a paradoxical relationship. Producers of intelligence want to be completely anonymous when sharing. However, it is the preference of CTI consumers to have proof that the intelligence they consume originates from a reputable source \cite{murdoch2015anonymity}. Moreover, the group of users who make up a sharing platform should have governance over the information shared, and consequently be able to hold users who share false information accountable. As a result, the way in which CTI sharing platforms manage privacy, trust, and accountability is an important challenge.  %{\color{red}(Shan: What platforms- mention it?)} {\color{blue}(KD: This isn't a platform specific thing. I am more saying that sharing platforms have to consider how they manage these factors. Reworded to hopefully clarify this.)}

%{\color{red}(Shan: One thing is missing in this discussion is the decentralisation aspects of trust, privacy, and accountability. Plz add one or two lines here)} 
%{\color{blue}(KD: I am not to sure what you mean here. I will email you about this.)}  {\color{red}(Shan: I wanted to say that a centralised system can give more trust privacy. But they are ineffective in IoT. So we need decentralised trust and privacy. Then how do we manage the accountability? What would be the trade-off?)}

\section{Opportunities}
\label{opportunities}

\begin{figure*}
\centering
\includegraphics[scale=0.49]{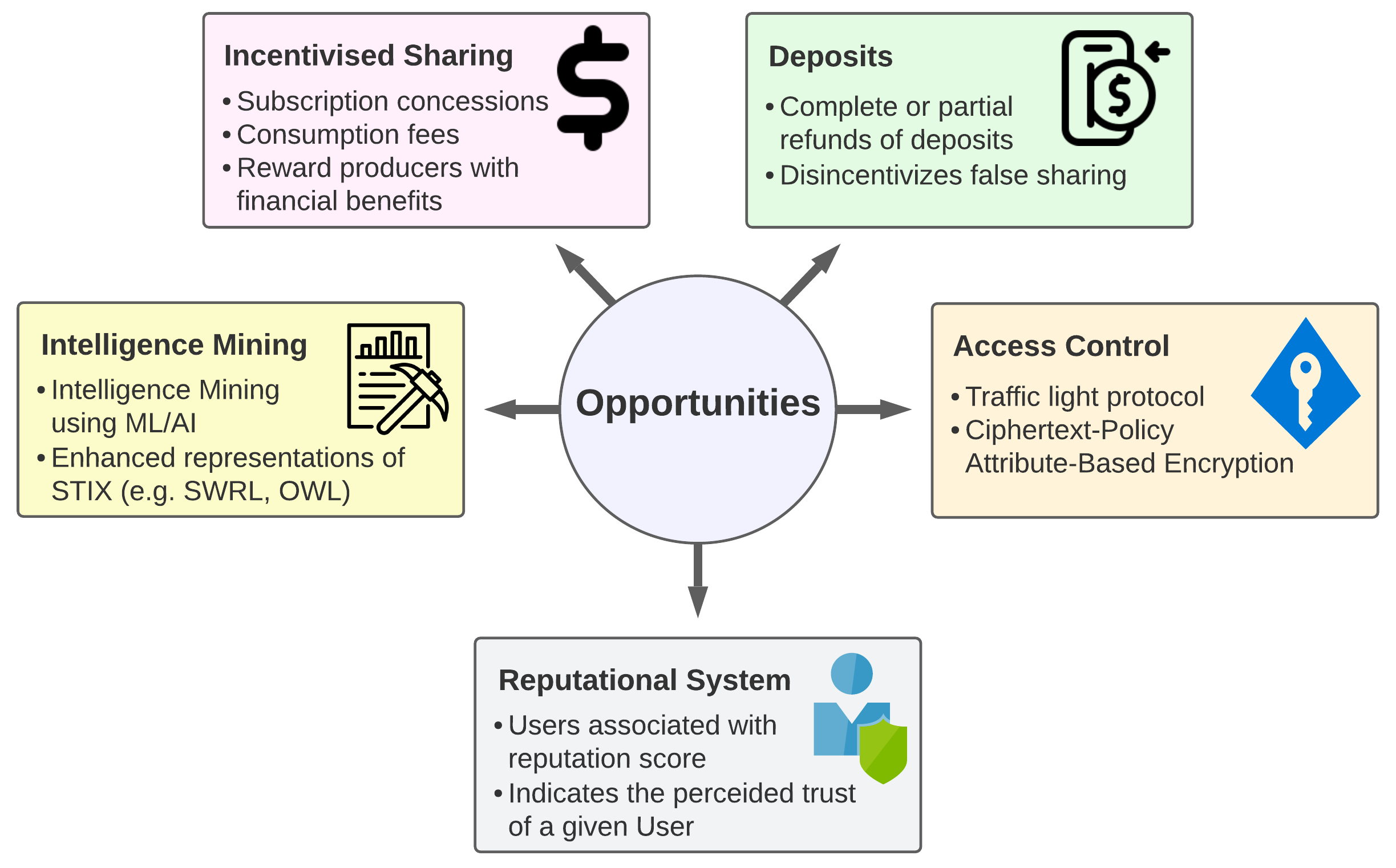}
\caption{Blockchain-based CTI sharing opportunities.}
\label{fig:x opportunities}
\end{figure*}

In this section, we discuss a list of opportunities (cf. Fig.~\ref{fig:x opportunities}) for blockchain-based CTI sharing. These opportunities aim to highlight how the characteristics of blockchain can be leveraged to provide novel solutions to the challenges discussed in Section~\ref{challenges}.   

%DONE
\subsection{Incentivised Sharing}
To help alleviate the producer consumer imbalance discussed in Section~\ref{pc-imbalance}, several incentive schemes can be implemented. In this section we will discuss two examples that illustrate how incentivised sharing can be achieved using blockchain. %{\color{green} but you have provided three examples.} {\color{blue} (KD: DEALER is only an example of Consumption Fees. I removed its heading to remove this confusion)}

\subsubsection{Concessions:}
Some blockchain-based sharing platform, such as \cite{nguyen2021blockchain}, use subscription fees to create permissioned sharing groups. Consequently, users are required to pay an authority a subscription fee to participate, consume and or share CTI, for a given time period. To incentivise users to share CTI and not just consume it, concessions can be given to users who contribute intelligence. As a users contributions are stored using blockchain (e.g. In a Smart Contract or directly on-chain), an auditable and immutable record of these transactions is maintained. This record can thereafter be used by an authority to determine a users subscription fee once their previous subscription has expired. In the case were a users record shows that they have shared valuable intelligence, the price of their next subscription can be lowered to incentives them to continue making valuable contributions going forward.    

An example of a sharing model that implements concession based incentives is \cite{nguyen2021blockchain}. In this model, the authors use subscription discounts to reward CTI producers for their contributions. As part of their implementation, proposal \cite{nguyen2021blockchain} provides a CTI producer with a discount each time they share intelligence that is considered high-quality by a set of verifiers. This design therefore allows users who continually share high-quality CTI to significantly reduce their subscription fees. To achieve this, CTI sharing is completed using the following steps; 

\begin{enumerate}
    \item CTI producer adds CTI to the blockchain.
    \item Three random verifiers are selected from a trusted group.
    \item Verifiers rate the CTI's quality using pre-determined metrics.
    \item If the majority of verifiers rate the given CTI as high-quality then both the producer and verifiers are given a discount on their next subscription.
    \item If the majority of verifiers rate the given CTI as low-quality then only the verifiers are given a discount on their next subscription.
\end{enumerate}

\subsubsection{Fees:} 
Another example of how blockchain can be used to combat free riding behavior within CTI sharing is consumption fees. Unlike subscription concessions, consumption fees require consumers to pay producers to access CTI that they have shared \cite{riesco2020cybersecurity} \cite{menges2021dealer}. In essence, consumption fees aim to create a market place where CTI can be exchanged between organisations for currency. Due to the trustless properties of blockchain, CTI can be exchanged between two organisations without the need to pre-establish trust or use a third party. Instead, self manged Smart Contract can be used to facilitate the exchange of CTI and cryptocurrency between two organisations. By creating a blockchain-based CTI marketplace, producers who actively share valuable CTI have the ability to profit significantly from doing so. 

Two main approaches can be used to implement consumption fees within blockchain-based architectures. 
\begin{enumerate}
    \item \textbf{Standard fee:} A predefined fee is payed to producers when other organisations consume their CTI \cite{riesco2020cybersecurity}.
    
    \item \textbf{User defined fee:} Producers specify a consumption fee which is payed each time a user access it \cite{menges2021dealer}. Can be implemented as a producer defined parameter in a Smart Contract.  
\end{enumerate}

Decentralised incEntives for threAt inteLligEnce Reporting and exchange (DEALER), is an example of a blockchain-based CTI sharing platform that implements a user defined consumption fee to incentivise CTI sharing \cite{menges2021dealer}. The below steps summarise how DEALER implementes user defined consumption fees.

\begin{enumerate}
    \item CTI producer adds CTI to the blockchain. During this process, the producer can define a sale price. If a producer does specify a sale price, they also have to pay a verification fee.
    \item In the case where a sale price is specified, three trusted verifiers review the associated intelligence using pre-determined metrics. 
    \item The results of each verifiers analysis are added to the blockchain to indicate to buyers the quality of the given intelligence. Moreover, each verifier is given a proportion of the verification fee.
    \item When a buyer purchases some intelligence they are required to pay the associated consumption fee, if specified by the producer.  
\end{enumerate}

As discussed in Section~\ref{pc-imbalance}, an imbalance between the producer and consumers roles exists within CTI sharing. Consequently, it is critical that sharing platforms seek to address this imbalance by providing producers with more direct benefits. In this section we highlighted a number of way in which blockchain-based sharing platforms can implement different incentive schemes to combat the effects of the producer consumer imbalance.        

\subsection{Deposits}
\label{deposits}
In Section~\ref{data-validity} the issue of false sharing was discussed. To disincentivise CTI producers from participating in this behaviour negative financial punishments can be used. In the case of blockchain-based platforms, existing technologies that support the exchange of cryptocurrency can be utilised for this purposes (e.g. Ethereum). Moreover, many of these platforms also allow self managed Smart Contracts to exchange cryptocurrency autonomously, thus removing the need for a trusted third party \cite{swan2015blockchain}. As a result, Smart Contracts can be utilised to implement conditionally refundable deposits in a trustless, auditable and verifiable manner.  

Conditionally refundable deposits can be utilised by blockchain-based CTI sharing platforms to introduce negative financial punishments for CTI producers that participate in false sharing. In this case, when a producer shares some intelligence they could be required to pay a deposit, some amount of cryptocurrnecy, to a Smart Contract. Once payed, a consensus algorithm defined within the Smart Contract can be used to verify the integrity of the shared intelligence \cite{gong2020blocis}. Given that this verification process occurs on-chain, the results are immutable and transparent to both the original producer as well as future consumers. Furthermore, the autonomous and deterministic nature of Smart Contracts allows them to hold cryptocurrency in escrow without the need for pre-established trust.    

In the case where shared intelligence is found to be credible, the initial deposit can be payed back either in full or partially to the original producer automatically by the Smart Contract. On the other hand, when false sharing is found to have occurred this deposit can either be held by the Smart Contract, burned or distributed to users involved in the verification process \cite{swan2015blockchain}. By punishing users who participate in false sharing, persistent efforts to do so on a large scale are deterred due to the associated financial cost. %{\color{red}(Shan: Cost of what?)} {\color{blue} (KD: Done)}.  

\subsubsection{BLOCIS:}
In \cite{gong2020blocis}, the authors use conditionally refundable deposit to disincentivise stakeholders from deliberately sharing false/incorrect CTI. When a registered BLOCIS user shares CTI, they use a pre-defined Data Report Contract (DRC). This contract takes as input the given CTI as well as a deposit. Once added to a specific feed, an evaluation function ($\pi$) is used to assess the validity of the reported intelligence. The novelty that BLOCIS proposes is that $\pi$ takes as input both the reputational score of the producer as well as their deposit. If the output of $\pi$ indicates the given CTI is false, the deposit is not refunded back to the producer. When simulated in a test environment, the BLOCIS model was found to successfully disincentivise users who made malicious contributions. Fig. 5 in \cite{gong2020blocis} demonstrates both the financial and reputational damage that users who participated in false sharing suffered over an extended period of time.

%{\color{green} Zahra: Are you refereing to a figure in another paper? If you need the figure, you can make your own version here or briefly explain it instead.} {\color{red} (Shan: As we discussed put one two lines describing the figure from the other paper)}

\subsubsection{Considerations: }
While deposit-based disincentive schemes are focused on punishing malicious producers, considerations must also be made to ensure honest producers are not deterred from sharing. Although the self managed nature of Smart Contracts can provide producers with a trustless way to exchange cryptocurrency, factors such as the amount of currency and consensus used to determine if a contribution is false must be considered. For example, if producers are required to pay a constant amount of cryptocurrency, the extremely volatile nature of currency markets could cause producers not to share at particular times \cite{sams2015note}. Moreover, if consensus methods are dependent on validation of intelligence from a set of validators, then they themselves could become by subject to malicious attacks. Given cryptocurrency is at stake, we argue that malicious attacks could seek to compromise a subset of validators to deny the authentication of any intelligence. Lastly, if validators are directly incentivised through partial payment of deposits from intelligence deemed malicious, then validators might be more likely to classify honest contributions as malicious. All of these factors need to be considered carefully when designing a deposit-based disincentive scheme as they have the potential to affect honest producers as well. 

Aside from considerations related to how deposit-based Smart Contracts are designed, the method used to validate intelligence is another important factor. Fundamental to the success of conditionally refundable deposits is the ability of a verifier or group of verifiers to determine the credibility of CTI. However, currently a method that deterministically classifies CTI as false is considered an open challenge \cite{gong2020blocis}. Consequently, platforms that implement disincentive schemes are likely to encounter cases where CTI is wrongly considered malicious and an honest producer is punished.

\subsection{Reputational System}
Another way blockchain-based solutions can mitigate against false sharing is use of reputational systems. Unlike deposits, reputational systems do not punish malicious users monetarily. Instead, they associate each user with a reputation score (e.g. 1-100) that represents their perceived trustworthiness.

In the context of CTI sharing, a users reputational score can be used to directly affect their ability to consume and or share intelligence within a group \cite{xiaohui2021reputation} \cite{gong2020blocis}. For example, if a CTI producer shares some CTI, their associated reputational score could be used to indicate to validators and or consumers the level to which they should trust it\cite{gong2020blocis}. As a result, intelligence shared by a users with a relatively low reptuational score might be subject to more thorough inspection by validators. In the opposite case, users who have a relatively high reputational score, may be subject to less thorough inspection by validators. Furthermore, these highly trusted users might be able to consume more sensitive intelligence that might otherwise have been unavailable to them. 

A successful reputational system has the potential to stop a user or group of users from continually false sharing \cite{gong2020blocis}. Given that a users reputational score is tightly coupled with their perceived trust, efforts to continually false share can be predicted to become harder over time. 

\subsubsection{Proof-of-Reputation (PoR)}
is a blockchain-based consensus algorithm that was proposed by \cite{xiaohui2021reputation} specifically for CTI sharing. In their model, each node in the network has an associated reputational score between 1 and 100. Fundamentally, this score seeks to capture how trustworthy a user is based on the creditably of their previous contributions. Importantly, all of the actions taken by a node (e.g. Voting, Sharing CTI) influence its reputational score over time.  

When an organisation shares CTI, other nodes on the network calculate a reputation value which is used to judge if it should be added to the blockchain. The results of this reputation-based consensus are used alongside more traditional validation methods to try and mitigate against false sharing. Moreover, a contributing nodes reputational score is adjusted over time based on the results of this process. Critical to the integrity of this process is a predefined trust threshold. This trust threshold defines the point at which a node is considered trustworthy. As a result, if a nodes score drops below this threshold, then it is considered untrustworthy and cannot participate further. \\

The above PoR consensus algorithm exemplifies how the inherent properties of blockchain discussed in Section \ref{blockchain}, can be utilised to facilitate reputational systems without the need for a trusted third party. In particular the immutable, transparent and auditable properties of blockchain allow each node to calculate the reputational scores of others, thus removing the need for a centralised authority. Similar to \cite{xiaohui2021reputation}, the BLOCIS architecture proposed by \cite{gong2020blocis} manages reputational scores with self managed Smart Contracts. Like deposits, these Smart Contracts contain a predefined consensus algorithm that can be used to manage the reputational scores of each user over time in a trustless way. 

As mentioned in the in Section \ref{deposits}, the ability to deterministically validate CTI is still an open challenge. Given that reputational systems require a verifier or group of verifiers to determine the credibility of CTI, their success is dependent on the accuracy of the validation method used.

\subsection{Access Control}
Blockchain-based sharing platforms can use several methods to provide producers with control over who has access to the intelligence they share. Access control in this case, refers not just to the ability of CTI producers to control who has access to their intelligence, but also in what way~\cite{pal2020access} \cite{pal2018fine} \cite{pal2018policy}. For example, a particular producer might want to share CTI with a small trusted group. However, they only want to disclose the specific attribute values (e.g., IP addresses) associated with it to one of the organisations.

While access control can be implemented by centralised architecture, blockchain is able to facilitate the fine grained access control required for CTI sharing in a trustless way. The following list outlines how a number of the key properties of blockchain can be leveraged to provide access control in a trustless way~\cite{pal2021internet-book}.  

\begin{itemize}
    \item \textit{Decentralised:} As a single authority does not control access based on a producers request, greater integrity is achieved. This means that producers have greater confidence that the control policy they define will be followed given its execution is not dependent on a centralised system. 
    \item \textit{Immutability:} CTI producers can be confident that the access control parameters they define cannot be altered by another user for their benefit.
    \item \textit{Smart Contracts:} Provides a framework to allow stakeholders to define the access control for the intelligence they share. Moreover, the self managed nature of Smart Contracts ensure that these access control policies are executed autonomously.    
\end{itemize}

\subsubsection{Traffic Light Protocol (TLP)} is an example of an access control method that can be implemented as part of a blockchain-based sharing platform \cite{nguyen2021blockchain}. TLP defines a robust access control structure that gives producers the ability to specify who CTI is shared with. This is achieved by allowing producers to specify a sharing level from a predefined list. Each of these predefined sharing levels is simply a control policy that specifies which users can access the CTI. Table \ref{table:TLP} is an example of how a TLP policy could be structured.

\newcolumntype{P}[1]{>{\centering\arraybackslash}p{#1}}
\begin{table}[ht]
    \setlength{\extrarowheight}{5pt}
    \centering
    \caption{Example of a TLP implementation by \cite{nguyen2021blockchain} \cite{homan2019new}}
    \small
    \begin{tabular}{|P{3cm}|P{7.8cm}|}
        \hline
        \textbf{Channel} & \textbf{Description}  \\
        \hline
        Red & Private channel between two stakeholders only \\
        \hline
        Orange & Disclosure to only a certain group of stakeholders defined by the CTI producer. \\
        \hline
        Green & Disclosure to an entire group of stakeholders. In the case of private blockchain this is restricted to anyone who has access to it. \\
        \hline
        White & Public disclosure which is accessible to anyone. \\
        \hline
    \end{tabular}
    
    \label{table:TLP}
    %\scl
\end{table}

\subsubsection{Ciphertext-Policy Attribute-Based Encryption (CP-ABE)}
is another method that can be used to give producers with fine grained access control \cite{preuveneers2020distributed}. In the case of CP-ABE, when a producer shares CTI, they encrypt it using attribute-based encryption methods \cite{bethencourt2007ciphertext}. The ciphertext that results from this process is then added to the blockchain. When a user access this ciphertext they are able to decrypt parts of it based on their own attributes. As a result, users can define highly specific fine grained access control policies using CP-ABE.  

For example, a CP-ABE policy might require that an organisation is a \textit{ICS-ISAC} member to view a subset of the CTI. Furthermore, it might also specify that only a specific subset of these organisations can access the specific details of the hardware affected by a ransomware attack. This example demonstrates how CP-ABE can be used to construct fine grained access control policies specific to a producers needs. \\

Both TLP and CP-ABE are examples of access control methods that can be implemented using blockchain. Importantly, these methods provide CTI producers with better control over who consumes the intelligence they share in a trustless way. In Section~\ref{privacy}, the issue of privacy was discussed. During this discussion, it was highlighted that fear of reputational damage was a significant barrier that stopped some organisations from sharing. While greater access control does not provide a complete solution to this problem, we argue it has the potential to cause more organisations to share within closed groups given their privacy-preserving nature. Moreover, if key regulatory bodies are incorporated into sharing platforms, these frameworks can further help organisations meet their legal and regulatory obligations without having to use secondary sharing mechanisms \cite{nguyen2021blockchain}.     

\subsection{Intelligence Mining}
In Section \ref{intelligent-intelligence}, it is noted that not all types of CTI are equivalent in their ability to describe threats and subsequently be used to implement mitigation strategies against them. Given that the process of generating CTI is dependent on the capabilities of the sharing organisation, it cannot be expected that all organisations are capable of generating high-level intelligence. As a result, strategies to create high-level intelligence from aggregated sources of low-level intelligence have the potential to shift sharing towards more intelligent intelligence. Furthermore, this process also allows organisations which do not have the resources to generate high-level intelligence themselves to still contribute.

Intelligence mining can be defined as the process of deriving high-level intelligence from low-level intelligence already stored on the blockchain \cite{tounsi2018survey}. The immutable and auditable properties of blockchain are able to facilitate mining in a trustless way. Given that low-level intelligence used as part of the mining process is immutable and accessible by each organisation on the network, high-level intelligence that is derived from it can be validated by other organisations. As a result, the ability to mine high-level intelligence in a trustless way has the potential to allow blockchain-based CTI sharing platforms to provide participating organisations with more advanced threat mitigation.  

Proposal \cite{ riesco2020cybersecurity} provides an example of how STIX, Semantic Rule Language (SWRL) and Web Ontology Language (OWL) can be combined to create more meaningful and interpretable representations of CTI. The use of these tools together has great potential in the area of intelligence mining, as CTI represented in this way allows semantic reasoners to infer new knowledge \cite{ riesco2020cybersecurity}. Furthermore, extending traditional representations of CTI could also pave the way for Machine Learning (ML)/Artificial Intelligence (AI) approaches to intelligence mining. In \cite{ jadidi2020securing}, it was demonstrated that ML algorithms were able to generate CTI from a single organisations network logs stored using blockchain. Therefore, it could be possible to extend this approach further to generate more intelligent intelligence, from large amounts of aggregated CTI expressed using STIX, SWRL and OWL.  

\section{Related Work and Discussion}
\label{related-work}

%{\color{red} Shan: Remember page, word, references are not an issue, so we can write more. What is missing now in this paper, a related work section. Write 6-8 papers summary (around a page) of existing papers in this domain. Say what they did and why we need to further study this domain. For say, paper [A] say CTI but only for general IT systems and do not say blockchain. Paper [B] did not consider accountability. Paper [C] says ... but unlike our study, they do not present opportunities, etc.} {\color{blue} KD: Done. I only discussed papers that aimed to highlight specific CTI sharing challenges. I did not include any blockchain-based models. Please set me know if this needs to change.} {\color{red} (Shan: As it is a blockchain based paper you need to write a few limitation in present blockchain based CTI sharing proposals have. I have written a summary in the Introduction, see it and you expand these three papers here)} {\color{blue}(KD: Done)}

%\subsection{CTI Sharing}
In this section, we present some related works on CTI sharing and the integration of blockchain platforms for CTI sharing and provide a discussion on the findings of this chapter. Several studies discuss the importance of CTI sharing in information security and general computing systems \cite{wagner2019cyber} \cite{tounsi2018survey} \cite{kure2019cyber} \cite{gong2017barriers}. However, most of them discuss CTI sharing from the lens of traditional centralised computing approaches. Subsequently, few publications considering how blockchain-based approaches can overcome existing challenges are present in the literature. In this section, we aim to discuss the contributions of several publications that outline challenges associated with CTI sharing.  

Proposal \cite{wagner2019cyber}, provides a comprehensive insight into what CTI sharing is and how it is commonly performed. Furthermore, it also discusses a number of important CTI sharing concepts including -- what CTI is, how it can be shared, and most notably what benefits and risks are associated with sharing. Of particular note, the authors highlight the importance of privacy and anonymity in CTI sharing. However unlike \cite{wagner2019cyber}, in this chapter, we extended these ideas to consider the relationship between privacy, trust, and accountability.

In \cite{tounsi2018survey}, a survey on technical threat intelligence was conducted. Like \cite{wagner2019cyber}, \cite{tounsi2018survey} provides a good insight into the key concepts which define CTI sharing. This paper specifically seeks to provide a clear definition of what threat intelligence is and what some of the associated challenges in this space are. An important challenge highlighted by \cite{tounsi2018survey}, is intelligent intelligence. Moreover, their suggestion that big data analysis could be applied to threat intelligence was extended by our work to focus on how these concepts can be applied to blockchain specifically.  

Proposal \cite{sauerwein2017threat} provides a comprehensive study into the current challenges associated with CTI sharing platforms (e.g. MISP). As part of their research, they investigate twenty two sharing platforms and derived a list of eight key findings. A number of which are discussed in Section \ref{challenges}. While their research was mostly focused on centralised architectures, their insights into existing challenges allowed us to highlight how blockchain-based architectures can provide novel solutions to them. 

In \cite{abu2018cyber}, the authors perform a comprehensive literature review into the current use CTI. As part of their findings, they outline four main challenges of which three were discussed in this chapter. However, unlike our approach, this research does not explore how blockchain-based solutions can provide novel solutions to these challenges. 

%\subsection{Blockchain-based CTI Sharing}
Recently, a diverse range of blockchain-based CTI sharing models have been published. In this chapter, we discussed a number of novel features present within a subset of these models which we feel represent the current state-of-the-art.

We argue that \cite{riesco2020cybersecurity} currently presents the most comprehensive blockchain-based CTI sharing platform, as it addresses a number of the challenges presented in this chapter. As part of their model, the authors integrate a number of features which address the producer consumer imbalance, intelligence intelligence, and legal and regulatory factors. However, it must be noted that while this model does provide trust and accountability, it is achieved at the cost of privacy-preserving anonymity. 

DEALER is a blockchain-based CTI sharing platform presented by \cite{menges2021dealer}, which like \cite{riesco2020cybersecurity}, presents novel solutions to a number of the challenges discussed in this chapter. The DEALER proposal provides solutions to the producer consumer imbalance and legal and regulatory factors. Moreover, this proposal also integrates a quality assurance method which provides a heuristic approach to solving the challenge of data validity. It must be noted, however, that while a heuristic approach to the issue of data validity has the potential to be effective, it does not completely mitigate against false sharing.   

Few models present in the current literature provide a robust framework that balances privacy, trust, and accountability, as defined in Section \ref{privacy}. We argue that \cite{allouche2021trade} presents the most comprehensive approach to balancing these factors. The authors of this platform propose a framework which allows CTI producers to share intelligence semi-anonymously while still facilitating trust and accountability. However, the major limitation of this framework is that a single trusted authority has the ability to reveal the identity of any CTI producer, subsequently creating a single point of failure.   

We find there are various challenges in CTI sharing, and blockchain is a promising solution to gain opportunities in most cases. However, there is still a list of open research questions that need to be resolved. We list a few of them as follows:

\begin{itemize}
    \item How the properties of blockchain and other cryptographic constructs be used to create a blockchain-based CTI sharing model that provides a balance between privacy, trust, and accountability?
    
    \item How can shared CTI be deterministically validated to ensure false sharing is not possible?
    
    \item How can ML/AI be utilised along side current approaches (e.g. STIX, SWRL, OWL) to facilitate the sharing of more intelligent intelligence?
\end{itemize}

%Several studies discuss the importance of CTI sharing in information security and general computing systems. However, most of them are centralised and do not consider a variety of challenges, e.g., ..... . In recent years, the use of blockchain technology to enable CTI data among organisations has been getting much attention, where information to be shared and maintained in a decentralised and distributed network, in a trustless manner. For example, proposal [111] discuss a ..... However, unlike our approach, it does not present the critical challenges of blockchain integration for CTI sharing at scale. Proposal [222] present ....

%%%
\section{Conclusion}
\label{conclusion-future-work}
The drastic evolution of the threat landscape, brought about by the emergence of Internet of Things (IoT) technology, has caused organisations to find new ways to better manage their cyber risks. This appetite for tools that better mitigate against potential threats has driven the development for a number of Cyber Threat Intelligence (CTI) sharing platforms (e.g., MISP). In this chapter, we defined a number of general CTI sharing challenges including the producer consumer imbalance, legal and regulator factors, intelligent intelligence, data validity, and privacy, trust and accountability. These general CTI sharing challenges were then used to deliver a list of opportunities present within the blockchain-based space. These opportunities included deposits, access control, reputational systems, intelligence mining and incentivised sharing. Finally, we explored several existing proposals and determine a list of unique future research questions for efficient and secure CTI sharing using blockchain. 

%{\color{green} As the future gaps were discussed in the previous chapter, this title should only be "conclusion"}{\color{blue} (KD: Done)}

%\subsection{Future Research Questions}
%\label{future-research-questions}

%%%\bibliographystyle{splncs04}
\bibliographystyle{IEEEtran}
\bibliography{bare_jrnl}
%

% \end{thebibliography}

\end{document}